\documentstyle[twocolumn,aps,prb,epsf]{revtex}
\input epsf
\begin{document}
\draft
\title{Low temperature structural model of hcp solid C$_{70}$}

\author{G. Ghosh, V. S. Sastry and T. S. Radhakrishnan}

\address{Materials Science Division, Indira Gandhi Centre for Atomic
Research, Kalpakkam 603 102, India.}

\address{%
\begin{minipage}[t]{6.0in}
\begin{abstract}
	We report intermolecular potential energy calculation for solid
C$_{70}$ and determine the optimum static orientations of molecules at low
temperature which are consistent with the monoclinic structural model proposed
by us in an earlier report [Solid State Commun., {\bf 105}, 247 (1997)].  This
model indicates that the C$_5$-axis of the molecule is tilted by an angle
$\sim$ 18$^o$ from the monoclinic b-axis in contrast with the molecular
orientation proposed by Verheijen et al.[J. Chem.  Phys., {\bf 166}, 287
(1992)] where the C$_5$-axis is parallel to the monoclinic b-axis. In this
calculation we have incorporated the effective bond charge Coulomb potential
together with the Lennard-Jones potential between the molecule at the origin of
the monoclinic unit cell and its six nearest neighbours, three above and three
below. The minimum energy configuration for the molecular orientations
turns
out to be at $\theta$ = 18$^o$, $\phi$ = 8$^o$ and $\psi$ = 5$^o$ where
$\theta$, $\phi$ and $\psi$ define the molecular orientations.
\end{abstract}
\end{minipage}}
\maketitle
\newpage

	The orientational ordering transformation in solid C$_{60}$ is well
understood\cite{HEI}. Potential energy calculations using only the
Lennard-Jones potential derive the low temperature structure of solid C$_{60}$
to be orthorhombic\cite{GUO}, in sharp contradiction with the experimentally
observed simple cubic structure. Lu et al.\cite{LU} could successfully derive
the simple cubic structure for the low temperature phase of solid C$_{60}$ by
adding a Coulomb potential term to the Lennard-Jones potential. This Coulomb
interaction between neighbouring molecules arises from the effective charges at
the single bonds and the double bonds of the molecules in the ordered phase at
low temperature.  In the case of solid C$_{70}$, the elongated molecular shape
causes much greater diversity in the structural phase transformation sequence
as well as in the orientational ordering of molecules. One of the first
experimental studies of structural phase transformation, on cooling, in solid
C$_{70}$ is by Verheijen et al.\cite{VER} and they reported the following
sequence of phase transformations

fcc $\rightleftharpoons$ rh $\leftarrow$ hcp-1 (c/a $\approx$ 1.63)  $\rightleftharpoons$ hcp-2 (c/a $\approx$ 1.82) $\rightleftharpoons$ monoclinic 

where the fcc is the high temperature phase and the monoclinic is the low
temperature phase. The lattice parameters reported for the monoclinic unit cell
are, a = c = 19.96 $\AA~$, b = 18.51 $\AA~$ and $\beta$ $\approx$ 120$^o$ with space
group {\sl P112$_1$/m}. They proposed a structural model for this monoclinic
structure in which the orientation of the long axis (C$_5$) of the static
molecules is parallel to the monoclinic b-axis (i.e., the original hcp c-axis).
Subsequently, they have also carried out the potential energy calculation using
the atom-atom Lennard-Jones potential and the corresponding result was in
agreement with their model.  Later, Que et al.\cite{QUE} pointed out that the
unique angle $\beta$ for a monoclinic structure with space group {\sl
P112$_1$/m} cannot be 120$^o$, but should be less than 120$^o$ which is not
consistent with the molecular orientations proposed by Verheijen et
al. They have carried out a detailed group theoretical analysis
of librations in solid C$_{70}$ and proposed an orthorhombic structure for
which the two space groups {\sl Pbnm} or {\sl Pnma} have the same Raman
modes. Nelissen et al.\cite{NEL} have carried out a harmonic lattice
dynamical calculation using the Lennard-Jones potential which proposed
that the structure of solid C$_{70}$ should be monoclinic at low
temperature with space group {\sl P2$_1$/m}. This structure requires two 
independent angles which they defined as $\phi1$ and $\phi2$ and pointed 
out that for their structure $\phi1$ + $\phi2$ should be 36$^o$. Later,
Agterberg
et al.\cite{AGT1} have pointed out that the calculation of Nelissen et
al. can also admit of an orthorhombic structure with space group {\sl
Pbnm}. In another detailed study using the anisotropic interaction
potential between two C$_{70}$ molecules in terms of Cayley-Klein
parameters, they have confirmed that the structure could be
orthorhombic with space group {\sl Pbcm} or {\sl Pbnm}\cite{AGT2}. The
experimental results of van Smaalen et al.\cite{SMA} also conclude in
favour of an orthorhombic structure with space group {\sl Pbnm} of the
solid
C$_{70}$ at low temperature. It is to be noted that this structure has
resulted by a two step transformation from the ideal h.c.p. phase at high
temperature. Oh et al.\cite{OH} have reported detailed
potential energy calculations using Girifalco-Lennard-Jones potential for
different molecular orientations depicting various possible structures of the
solid and the corresponding lattice parameters. They have compared their
results with experimental reports by others\cite{OH}. They reported two
possible transformation sequences on cooling, {\it viz.},

fcc $\rightleftharpoons$ ortho(I,II) $\rightleftharpoons$ rh $\leftarrow$ hcp I $\rightleftharpoons$ mono \\

or, fcc $\leftarrow$ hcp II $\rightleftharpoons$ hcp (III, IV) $\rightleftharpoons$ mono 

where the fcc is the high temperature phase and the monoclinic is the low
temperature phase. Here hcp I - IV correspond to the c/a ratios of the unit
cells which are 1.84, 1.61, 1.50 and 1.53 respectively.  The ortho I and II
correspond to the lattice parameters, a = 16.2$\AA~$, b = c = 14.0$\AA~$  and a =
16.0$\AA~$ , b = 14.4$\AA~$, c = 14.8$\AA~$  respectively. The monoclinic structures in
both cases are the same and correspond to the C$_5$-axis of orientation being
parallel to the b-axis of the unit cell. This is similar to the molecular
orientation in the monoclinic structure predicted by Verheijen et
al.\cite{VER}, but the lattice parameters in the two reports are significantly
different. Pickholz et al.\cite{PICK}, while using atom-atom Lennard-Jones
interaction in their calculation, obtain the low temperature structure to be
hcp with c/a $\sim$ 1.77, instead of a monoclinic structure. The molecular
orientations in their model are, however, parallel to the unit cell c-axis.
Sprik et al.\cite{SPR1} have carried out constant pressure molecular dynamics
calculations incorporating Lennard-Jones potential together with the bond
charge Coulomb potential in a phase transformation study of  fcc solid
C$_{70}$. Their result indicates a two step transformation, {\it viz.},
fcc $\rightleftharpoons$ rh $\rightleftharpoons$ monoclinic 
which is in excellent agreement with the experimental results reported by
Christides et al.\cite{CHR}. In the low temperature monoclinic phase predicted
by these authors, the molecules in alternate layers of the structure are
oriented parallel to the (111) direction and to the (110) direction
respectively. We have reported in our earlier paper\cite{GG1}  that the ideal
hcp (c/a $\approx$ 1.63) solid C$_{70}$ undergoes a single step transformation,
on cooling, to a monoclinic phase which is in contrast with the two step 
transformation reported by
Verheijen et al.\cite{VER}.  We conjectured that the single step transformation
implies simultaneous freezing of the molecular rotations around their long axes
(C$_5$-axes) and the short axes (C$_2$-axes), as compared to sequential
freezing, which results in a two step transformation. The former is likely to
lead to a different monoclinic structure as compared to the
latter\cite{VER,QUE}. We had proposed a structural model, consistent with the
lattice parameters for the monoclinic structure measured from our x-ray
diffractogram (XRD) at 100 K, wherein the molecular C$_5$-axis has a tilt of
about 18$^o$ from the monoclinic b$_m$-axis. Our recent experimental studies
on the kinetics of phase transformation in hcp solid C$_{70}$\cite{GG2}
show that, except for very slow cooling rates, a two step transformation invariably
results. Such a two-step transformation may be similar to that reported by
Verheijen et al.\cite{VER} and Christides et al.\cite{CHR}. The low temperature
monoclinic structures and the C$_{70}$ molecular orientations reported by these
authors must, therefore, correspond to some local potential energy minima of the
system raised above the global minimum by misorientation stresses.

In this paper, we report the results of the potential energy calculation
pertaining to the monoclinic structure
obtained by us from our 100 K XRD. Since the cooling rate was very slow
(typically, 0.0033 K /min.) the monoclinic structure determined from our
experiment is expected to be close to the equilibrium structure at low
temperature. In this calculation, we have considered the interactions between
the seven molecules, one at the origin (0,0,0) of the monoclinic unit cell
and the remaining six at the six nearest neighbour positions, three above
and three below the basal plane. The orientation of the molecules has been
defined using the angles, $\theta$, $\phi$ and $\psi$, as explained later. The
final orientation of the molecules, as obtained by locating the minimum of the
potential energy has $\theta$ = 18$^o$, $\phi$ = 8$^o$ and $\psi$ =
5$^o$.

The structure of our C$_{70}$ sample at room temperature is ideal hcp\cite{GVN}
and the XRD pattern of this structure has been analysed in detail in an earlier
work\cite{MCV}. The hcp lattice constants are, a$_h$ = b$_h$ = 10.53$\AA~$, c$_h$
= 17.26$\AA~$ and $\gamma_h$ = 120$^o$.  The 100 K pattern fits well to a
monoclinic structure with lattice parameters, a$_m$ = 10.99$\AA~$, b$_m$ =
16.16$\AA~$, c$_m$ = 9.85$\AA~$ and $\beta_m$ = 107.75$^o$ and space group {\sl
P2$_1$ /m}\cite{GG1}.  It is clear that the following correspondence of lattice
parameters is appropriate in going from the hcp to the monoclinic: c$_h$
shrinks and forms b$_m$, b$_h$ increases and forms a$_m$, and a$_h$ shrinks to
form c$_m$.  This correspondence is shown in Fig.1. The hexagonal angle
$\gamma_h$ can be seen to reduce to the monoclinic unique angle $\beta_m$ by a
shear along the direction shown by the arrows in Fig.1(b).  It can be seen that
the above correspondence indicates a tilt of the C$_5$-axis of C$_{70}$
molecules by an angle of approximately 18$^o$ from the b$_m$-axis of the
monoclinic unit cell (see Fig. 1(b)).

In this calculation we have used the Lennard-Jones potential together with the
bond charge Coulomb potential. Following Lu et al.\cite{LU} in the ordered state 
of the solid there is less charge density distribution around the
C-C single bonds (long bonds) compared to that around the C-C double bonds
(short bonds). Such a charge density difference on the surface of the molecule
leads to short range intermolecular Coulomb interaction which decays as 1/
R$^5$ at long distances in the case of C$_{70}$.  In this calculation we have
used the same notation for the interaction sites as was done by Sprik et
al.\cite{SPR1} in their calculation for the fcc solid C$_{70}$. The
Lennard-Jones centres are taken to be the carbon atoms sites (C sites:
$\sigma_{CC}$ = 3.4$\AA~$) as well as the mid-points of the electron-rich C-C
double bonds (D sites: $\sigma_{DD}$ = 3.6$\AA~$, whereas $\sigma_{CD}$ =
3.5$\AA~$) and the magnitude of the interaction has been taken as,
$\epsilon$ = 2.964 meV\cite{GAU,ENC}.

The effective charge around the single bonds are assumed to be localised around
the C sites\cite{SPR1}.  Though in C$_{70}$ molecules, there are five types of
C sites, for the sake of simplicity all C sites can be treated to be of the
same type as in the case of C$_{60}$. If a C site carries an effective charge
q$_C$, the corresponding value at the D sites will be q$_D$ = - 2q$_C$. The
values of these effective charges are fractions of the electronic charge, e.g.,
q$_C$ = 0.175e and q$_D$ = - 0.35e\cite{SPR1}. The bonds around the equatorial 
region of the molecule
have a length intermediate between the length of a single bond and a double
bond and, therefore, are termed as intermediate bonds (or, I bonds). For these
bonds, $\sigma_I$ = 3.5$\AA~$ and q$_I$ = 0.5q$_D$.  Since the polar cap of the
molecule consists of both long bonds and short bonds, the Coulomb interaction
is stronger at the two opposite poles of a molecule in the ordered phase.
Therefore the bond charge interaction between two molecules in adjacent planes
is dominated by the polar caps and can be termed "polar cap" interaction. The
coordinates of the 70 carbon atoms in the molecule were generated from data
available in published literature\cite{AND}. The total interaction potential
between any two C$_{70}$\cite{LU} molecules is given by
\begin{eqnarray}
\nonumber
V_{12} = & \sum_{{i,j}} & 4\epsilon\left[\left(\frac{\sigma_{xx}}{\vert{\bf
r_{1i} -
r_{2j}}\vert}\right)^{12} - \left(\frac{\sigma_{xx}}{\vert{\bf r_{1i} - r_{2j}}\vert}\right)^{6}\right] + \\
\nonumber
& \sum_{{m,n}} &\left(\frac{q_m q_n}{\vert{\bf b_{1m} - b_{2n}}\vert}\right)
\end{eqnarray}
where the subscript x indicates interaction sites, e.g., C, D, or, I, and
r$_{1i,2j}$ are the coordinates of the atoms, b$_{1m,2n}$  coordinates of  bond
centres and q$_{m,n}$ are effective bond charges (e.g., q$_C$, q$_D$ and
q$_I$).

Using the above interaction potentials between the seven nearest neighbour
molecules for the lattice parameters measured from our 100 K x-ray
diffractogram, we optimise the molecular orientations with respect to the unit
cell parameters by minimising the total energy. The molecular orientations are
defined using the polar coordinates $\theta$ and $\phi$, where $\theta$ is the
polar angle, i.e., the angle between the b$_m$-axis and the molecular long axis
and $\phi$ is the azimuthal angle. Further minimisation of
the potential energy has been performed by varying the angle ($\psi$) which
defines the body rotation (spin) of the molecules around their C$_5$-axes of
the six molecules above and below the basal plane relative to the molecule at
the origin of the monoclinic unit cell.

We calculate the potential energy per cluster for each configuration of the
seven molecules defined by the values of $\theta$ between 0$^o$ and 90$^o$,
$\phi$ between 0$^o$ and 180$^o$, and $\psi$ between 0$^o$ and 72$^o$. In
Fig.2, we show the variation of potential energy with respect to $\theta$,
$\phi$ and $\psi$.  Clearly, for $\phi$ = 8$^o$ and $\psi$ = 5$^o$ the
potential energy is minimum at $\theta$ = 18$^o$.  This result is consistent
with the model we proposed for the molecular orientation in the monoclinic unit
cell experimentally measured by us\cite{GG1}. The minimum potential energy per
cluster obtained for our model is -3.923 eV which is significantly lower than 
other calculations reported in literature\cite{VER,OH}. It is to be mentioned here that
there are two other minima with respect to $\theta$ at 54$^o$ and 88$^o$ (not
shown in the figure). Though these three minima have energy values comparable
to each other, the energy barriers between them are nearly 0.04 eV which is too
large to spontaneously flip the molecule from one minimum to the other at low 
temperatures (100 K). These multiple minima indicate that the system may
have degenerate states with respect to $\theta$, or, that they may even correspond
to different structures which can be obtained by applying some external driving
force, e.g., pressure etc. The degeneracy may arise as a consequence of the
repulsive interactions between the C-C bonds as indicated by Sprik et
al\cite{SPR2}. in their extended model. Since the present calculation is based
on the static orientations of molecules (molecular static calculation) the
study of the structural transformation or the change of molecular orientations
with temperature cannot be carried out by this method. A detailed molecular
dynamics calculation is needed for this purpose which is in progress. 
Nevertheless, we argue that the equilibrium low temperature structure will be
consistent with the molecular C$_5$-axis tilted away from the unit
cell long axis, because of the polar cap interaction, as proposed in the present
model and supported by the potential energy calculation, rather than being 
oriented parallel
to it\cite{VER,QUE}. We recall from the introductory section that the molecular
dynamics calculation of Sprik et al.\cite{SPR1} leads to two competing
orientations for the molecules. A pictorial view of the optimised static
orientation of the seven molecules for our monoclinic lattice parameters is
shown in Fig.3. This picture is generated using a public domain software
MOLDRAW.

In conclusion, our results of potential energy calculation using the
Lennard - Jones potential together with the bond charge Coulomb potential 
support the structural model proposed by us in our earlier paper\cite{GG1}. 
The calculation shows that the minimum potential energy state
appears with a tilt of the C$_5$-axis of the molecule with respect to the
monoclinic long axis (b$_m$-axis) by an angle of 18$^o$. 
Further minimisation of the interaction energy between the neighbouring
molecules has been carried out using an angle $\psi$ which defines the
rotational configuration of the molecule around its own C$_5$-axis. The
molecular orientations obtained from our calculation have an implication of
significant polar cap interaction between the nearest neighbour molecules in
adjacent planes. The bond charge interaction is thus seen to play a key role in
determining the equilibrium low temperature structure of solid C$_{70}$. Our
calculations show also that the structural model proposed by us corresponds to
the most stable low temperature structure of solid C$_{70}$ reported so far.

 \acknowledgements
	We gratefully acknowledge M. C. Valsakumar and C. S.  Sundar for
valuable discussions.  G. Ghosh is thankful to the Director, Indira Gandhi
Centre for Atomic Research, for the visiting scientist position enabling
this work
to be carried out.

\begin{figure}

\caption{(a) High temperature hcp structural model; C$_{70}$ molecules are
freely rotating mimicking spheres,    (b) Monoclinic  unit  cell  model
proposed for the lattice parameters obtained from XRD pattern at 100 K (Ref.
10).} 

\caption{Variation of the potential energy in eV/ cluster (a) with $\theta$,
(b) with $\phi$, and (c) with $\psi$}. 

\caption{Final orientational configurations of the 7 molecules of the unit cell
arising from the potential energy calculation; a$_m$, b$_m$, and c$_m$
represent the monoclinic cell parameters.} 
\end{figure}
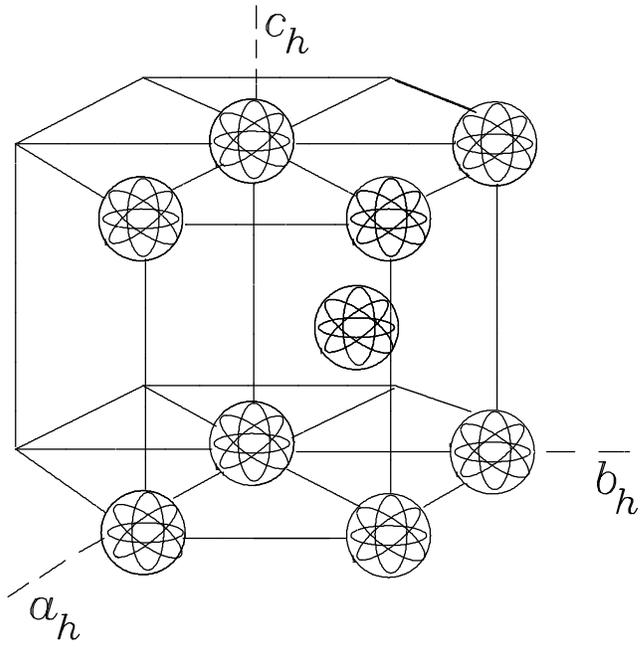
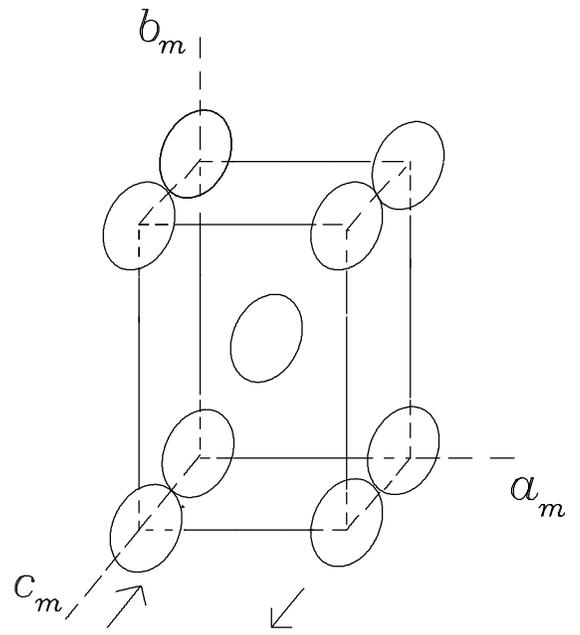
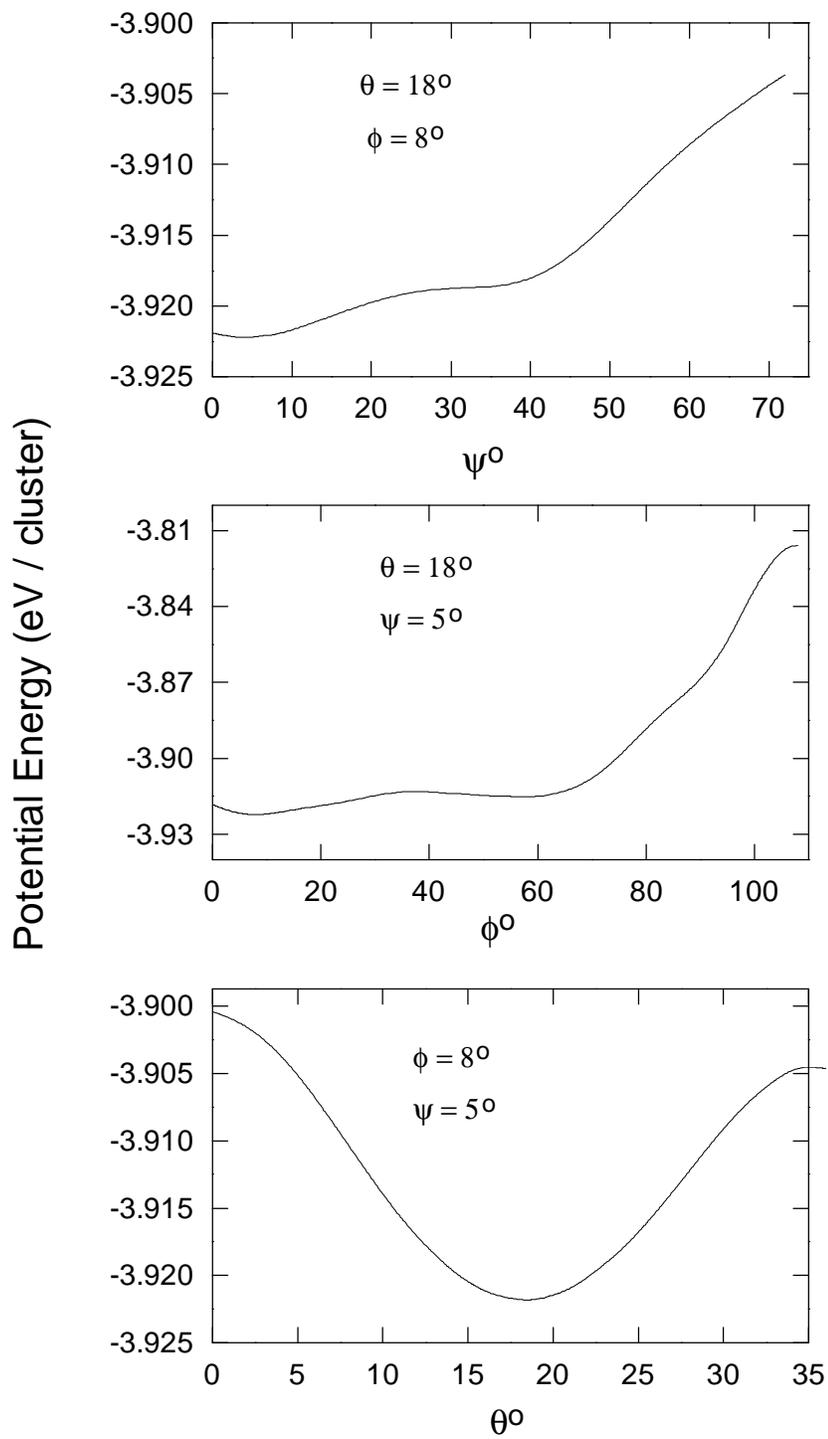
\end{document}